\documentclass[conference]{IEEEtran}
\IEEEoverridecommandlockouts
\usepackage{cite}
\usepackage{amsmath,amssymb,amsfonts}
\usepackage{graphicx}
\def\BibTeX{{\rm B\kern-.05em{\sc i\kern-.025em b}\kern-.08em
    T\kern-.1667em\lower.7ex\hbox{E}\kern-.125emX}}
    
\begin{document}
\title{Compiling Adiabatic Quantum Programs}
\author{\IEEEauthorblockN{1\textsuperscript{st} Faisal Shah Khan}
\IEEEauthorblockA{\textit{Center on Cyber Physical Systems} \\
\textit{and Department of Mathematics}\\
\textit{Khalifa University}\\
Abu Dhabi, UAE \\
faisal.khan@ku.ac.ae}
\and
\IEEEauthorblockN{2\textsuperscript{nd} Nada Elsokkary}
\IEEEauthorblockA{\textit{Department of Mathematics} \\
\textit{Khalifa University}\\
Abu Dhabi, UAE \\
sokkarynada@gmail.com}
\and
\IEEEauthorblockN{3\textsuperscript{rd} Travis S. Humble}
\IEEEauthorblockA{\textit{Quantum Computing Institute} \\
\textit{Oak Ridge National Lab}\\
Tennessee, USA \\
humblets@ornl.gov}
%\and
%\IEEEauthorblockN{4\textsuperscript{th} Given Name Surname}
%\IEEEauthorblockA{\textit{dept. name of organization (of Aff.)} \\
%\textit{name of organization (of Aff.)}\\
%City, Country \\
%email address}
%\and
%\IEEEauthorblockN{5\textsuperscript{th} Given Name Surname}
%\IEEEauthorblockA{\textit{dept. name of organization (of Aff.)} \\
%\textit{name of organization (of Aff.)}\\
%City, Country \\
%email address}
%\and
%\IEEEauthorblockN{6\textsuperscript{th} Given Name Surname}
%\IEEEauthorblockA{\textit{dept. name of organization (of Aff.)} \\
%\textit{name of organization (of Aff.)}\\
%City, Country \\
%email address}
}
\maketitle
\begin{abstract}
We develop a non-cooperative game-theoretic model for the problem of graph minor-embedding to show that optimal compiling of adiabatic quantum programs in the sense of Nash equilibrium is possible. 
% possible for certain graphs
\end{abstract}
\begin{IEEEkeywords}
Adiabatic quantum computing, non-cooperative game theory, minor-embedding, quantum compiling, network creation game
\end{IEEEkeywords}
\section{Introduction}
The adiabatic quantum computation (AQC) paradigm \cite{Farhi} is inspired by the physical processes known as quantum annealing where one starts with a quantum system in its lowest energy state and evolves the system ``slowly enough'' so that the new state is also that of lowest energy. More specifically, one initializes a quantum system so that its Hamiltonian $H_I$ is in the lowest energy state, and then interpolates to a final or problem quantum system with Hamiltonian $H_F$ via the {\it AQC program}
\begin{equation}\label{AQCprogram}
H(t)=A(t)H_I+B(t)H_F
\end{equation}
with $A(0)=1, B(0)=0$ and $A(T)=0, B(T)=1$, where $T$ is the run-time of the program $H(t)$. The value of $T$ determines if the solution to a problem obtained using an AQC program is worthwhile in comparison to run-times of programs on standard computers. The value of $T$ is determined inversely by the minimum spectral gap, which is the distance between the lowest energy state and the next excited state of the Hamiltonian $H(t)$. But this gap in turn may depend on how the program $H(t)$ is compiled for processing by an adiabatic quantum processor \cite{Klymko2014AdiabaticQP}. In the following sections, we consider the compilation of an AQC program as a type of network creation game \cite{Fabrikant} and show that an optimal compilation of AQC programs, in the sense of Nash equilibrium, is possible via fixed-point stability.
\section{Adiabatic quantum programming}

Compiling an AQC program $H(t)$ involves graph-theoretic considerations that are dependent on the hardware structure of the quantum processor. This is because the quantum hardware is effectively a graph, hereafter referred to as the {\it hardware graph} $\Gamma_H$, in which the vertices represent qubits and the edges represent interactions between the qubits. This specificity of the quantum hardware is based on the fact that both $H_I$ and $H_P$ are restricted in AQC to be of quadratic form, that is, 
\begin{equation}
\sum_{i \in V} \alpha_i X_i + \sum_{(i,j) \in E} \beta_{(i,j)}X_iX_j
\end{equation} 
where $E$ and $V$ are respectively the set  of edges and vertices of the {\it program graph}, $\Gamma_P$, that can be constructed from $H(t)$ by viewing it as a weighted adjacency matrix for $\Gamma_P$. Furthermore, $\alpha_i$ and $\beta_{(i,j)}$ are elements of the set $\left\{ 0,1\right\}$. 

Compiling an AQC program is the process of mapping $\Gamma_P$ into $\Gamma_H$, where the variables $X_i$ are taken to be the qubits in the hardware graph and the binary nature of $\alpha_i$ and $\beta_{(i,j)}$ is used to specify whether the qubits $X_i$ and $X_j$ are in use or not. In particular, the notion of minor-embedding \cite{Wagner} is fundamental to AQC program compilation, as we discuss furhter in the following section.

\subsection{Compiling AQC programs}

For the set of vertices and edges $V_P, E_P$ respectively of $\Gamma_P$, and the set of vertices and edges $V_H, E_H$ respectively of $\Gamma_H$, we make the following definitions. 
\\

\noindent {\bf Definition 1}: An {\it ideal compilation} is a function $f: V_P \rightarrow V_H$ such that if $(u,v) \in E_P$ then $(f(u),f(v)) \in E_H$. 
\\

\noindent {\bf Definition 2}: A {\it non-ideal compilation} is a relation $r: \Gamma_P \rightarrow \Gamma_H$ such that 
\\ \\
i) for $v \in V_P$, $r(v)$ is the vertex set of a connected subtree $T_v$ of $\Gamma_H$.
\\ \\
ii) for $(u,v) \in E_P$, there exists $i_u,i_v \in V_H$ such that $i_u \in T_u$, $i_v \in T_v$, and $(i_u,i_v) \in E_H$.
\\

An ideal compilation of an AQC program is obviously one for which the hardware graph has the smallest number of edges for the quantum processor to process. We consider this property to be a contribution toward saving processing time. There can be any number of non-ideal compilations of the program, but our goal here is to identify one which is optimal in the sense of Nash equilibrium in a network creation game given in \cite{Fabrikant}. The authors of this paper consider a given finite set of vertices as the set of players whose strategic choices are to identify an optimal subset of the players, so as to maximize connectivity between the players while minimizing the cost incurred in laying edges to establish the connectivity. We will consider a variation of this game in section \ref{megames}. First however, we review some essential game theory in the following section. 

\subsection{Non-cooperative Games and Nash Equilibrium}\label{AA}

We start by defining a non-cooperative game with finitely many player. 
\\ \\
{\bf Definition 3}: A $N$-player non-cooperative game is a function 
 \begin{equation}\label{Game}
G: \prod_{i=1}^N S_i \longrightarrow O,
\end{equation}

\noindent with the feature of non-identical preferences defined over the elements of the set of {\it outcomes} $O$, for every ``player'' of the game. The preferences are a pre-ordering of the elements of $O$, that is, for $l,m,n \in O$
\begin{equation}
m \preceq m,  \hspace{2mm}  {\rm and}  \hspace{2mm}  l \preceq m \hspace{2mm} {\rm and} \hspace{2mm}  m \preceq n \implies  l \preceq n.
\end{equation} 
where the symbol $\preceq$ denotes ``of less or equal preference''. 
\\

Preferences are typically quantified numerically for the ease of calculation of the payoffs. To this end, functions $G_i$ are introduced which act as the {\it payoff function} for each player $i$  and typically map elements of $O$ into the real numbers in a way that preserves the preferences of the players. That is, $\preceq$ is replaced with $\leq$ when analyzing the payoffs. The factor $S_i$ in the domain of $G$ is said to be the {\it strategy set} of player $i$, and a {\it play} of $G$ is an $n$-tuple of strategies, one per player, producing a payoff to each player in terms of his preferences over the elements of $O$ in the image of $\Gamma$.
\\ \\
{\bf Definition 4}: (Nash Equilibrium) \cite{Nash} A play of $G$ in which every player employs a strategy that is a best reply, with respects to his preferences over the outcomes, to the strategic choice of every other player. 
\\

In other words, unilateral deviation from a Nash equilibrium by any one player in the form of a different choice of strategy will produce an outcome which is less preferred by that player than before. Following Nash, we say that a play $p'$ of $G$ {\it counters} another play $p$ if $G_i(p') \geq G_i(p)$ for all players $i$, and that a self-countering play is an (Nash) equilibrium. 

Let $C_{p}$ denote the set of all the plays of $G$ that counter $p$. Denote $\prod_{i=1}^N S_i$ by $S$ for notational convenience, and note that $C_{p} \subset S$ and therefore $C_{p} \in 2^S$. Further note that the game $G$ can be factored as  
\begin{equation}\label{factor}
G:  S \xrightarrow {G_C} 2^S \xrightarrow{E} O
\end{equation}
where to any play $p$ the map $G_C$ associates its countering set $C_p$ via the payoff functions $G_i$. The set-valued map $G_C$ may be viewed as a preprocessing stage where players seek out a self-countering play, and if one is found, it is mapped to its corresponding outcome in $O$ by the function $E$. The condition for the existence of a self-countering play, and therefore of a Nash equilibrium, is that $G_C$ have a fixed point, that is, an element $p^* \in S$ such that $p^* \in G_C(p^*)=C_{p^*}$.

In a general set-theoretic setting for non-cooperative games, the map $G_C$ may not have a fixed point. Hence, not all non-cooperative games will have a Nash equilibrium. However, according to Nash's theorem, when the $S_i$ are finite and the game is extended to its {\it mixed} version, that is, the version in which randomization via probability distributions is allowed over the elements of all the $S_i$, as well as over the elements of $O$, then $G_C$ has at least one fixed point and therefore at least one Nash equilibrium. 

Formally, given a game $G$ with finite $S_i$ for all $i$, its mixed version is the product function 
\begin{equation}\label{mixedgame} 
\Lambda: \prod_{i=1}^N \Delta(S_i) \longrightarrow \Delta(O)
\end{equation}
where $\Delta(S_i)$ is the set of probability distributions over the $i^{\rm{th}}$ player's strategy set $S_i$, and the set $\Delta(O)$ is the set of probability distributions over the outcomes $O$. Payoffs are now calculated as {\it expected payoffs}, that is, weighted averages of the values of $G_i$, for each player $i$, with respect to probability distributions in $\Delta(O)$ that arise as the product of the plays of $\Lambda$. Denote the expected payoff to player $i$ by the function $\Lambda_i$. Also, note that $\Lambda$ restricts to $\Gamma$.

In these games, at least one Nash equilibrium play is guaranteed to exist as a fixed point of $\Lambda$ via Kakutani's fixed-point theorem.
\\ \\
{\bf Theorem 1}: (Kakutani fixed-point theorem) \cite{Kakutani} {\it Let $S \subset \mathbb{R}^n$ be nonempty, compact, and convex, and let $\Gamma: S\rightarrow 2^S$ be an upper semi-continuous set-valued mapping such that $\Gamma(s)$ is non-empty, closed, and convex for all $s \in S$. Then there exists some $s^* \in S$ such that $s^* \in \Gamma(s^*)$}.
\\

To be more specific, set $S=\prod_{i=1}^N \Delta(S_i)$. Then $S \subset \mathbb{R}^n$ and $S$ is non-empty, bounded, and closed because it is a finite product of finite non-empty sets. The set $S$ is also convex because it is the Cartesian product of the convex sets $\Delta(S_i)$. Next, let $C_p$ be the set of all plays of $\Lambda$ that counter the play $p$. Then $C_p$ is non-empty, closed, and convex. Further, $C_p \subset S$ and therefore $C_p \in 2^S$. Since $\Lambda$ is a game, it factors according to (\ref{factor})
\begin{equation}
\Lambda: S \xrightarrow{\Lambda_C} 2^S \xrightarrow{E_{\Pi}} \Delta(O)
\end{equation}
where the map $\Lambda_C$ associates a play to its countering set via the payoff functions $\Lambda_i$. Since $\Lambda_i$ are all continuous, $\Lambda_C$ is continuous. Further, $\Lambda_C(s)$ is non-empty, closed, and convex for all $s \in S$ (we invite the reader to check that the convexity of $C_p$ is immediate when the payoff functions are linear). Hence, Kakutani's theorem applies and there exists an $s^* \in S$ that counters itself, that is, $s^* \in \Lambda_C(s^*)$, and is therefore a Nash equilibrium. The function $E_{\Pi}$ simply maps $s^*$ to $\Delta(O)$ as the product probability distribution from which the Nash equilibrium expected payoff is computed for each player. 

\section{Non-ideal Compilation as a Non-cooperative Game}\label{megames}

We wish to interpret the non-ideal compilation problem as a type of network creation game, and use Kakutani fixed-point stability to show that a compilation at Nash equilibrium exists
Recall that in this game, the players choose any subset of vertices they wish to connect to. They can choose to build connections with all the players, with some of the players, or with none of them. However, if at the end of the game, a player is not connected to all the others, he/she is penalized. This is why even though it would cost a player to build a connection, in some cases, not building one could cost more.

In the compilation game on the other hand, a player's strategy is to choose a subset of vertices in $\Gamma_H$ such that a smallest possible subtree of $\Gamma_H$ spans this choice while keeping Definition 2 true. This is because the penalty to the players is defined in terms of the number of edges (which is directly proportional to the number of vertices) in the subtree it is mapped to.

Formally, the {\it non-ideal compilation game} is the function 
$$
C: S=\prod_{i=1}^n S_i \rightarrow 2^{S} \rightarrow \Gamma_H
$$
where the countering sets are calculated using the cost (payoff) function 
\begin{equation}
G_i(s) = \alpha \cdot |S_i|
\end{equation}
for player $i$, when strategy $s \in S_i$ is employed, and where the cost of laying down one edge is $\alpha$. Note that $G_i$ are linear and hence the countering sets will be convex. However, for Nash equilibrium guarantee via Kakutani's fixed-point theorem, we also need to ensure that the set of plays $S=(S_1,S_2, \dots, S_n)$ of the game is also compact and convex in the topological space that it resides in, that is, $\mathbb{R}^2$. 

Since each $S_i$ is a a tree inside $\Gamma_H \subset \mathbb{R}^2$, it is bounded; it is also closed in $\mathbb{R}$ with respect to the relative topology and hence closed in $\mathbb{R}^2$. Therefore, each $S_i$ is compact in $\mathbb{R}^2$ and hence so is $S=\prod_{i=1}^n S_i$. The set $S$ is convex only if each $S_i$ is, or in other words, each $S_i$ is a line segment or a single edge tree, in which case the Kakutani fixed-point theorem applies and a Nash equilibrium exists in the game $C$.

To consider a larger class of pure strategies in the game $C$ for which Nash equilibrium may be guaranteed, extend $S$ to its convex hull ${\rm Conv}(S)$ which is compact by Caratheodory's theorem \cite{Cath}. We now have fixed points for set-valued functions 
$$
F: {\rm Conv}(S) \rightarrow 2^{{\rm Conv}(S)}.
$$ 
It is a basic fact in topology that any compact, convex subset of $\mathbb{R}^m$ is homeomorphic to a closed ball $\mathbb{B}^m$ for all $m$. We therefore have that ${\rm Conv}(S) \cong \mathbb{B}^2$ and 
$$
F: \mathbb{B}^2 \rightarrow 2^{\mathbb{B}^2}.
$$

To get fixed points for functions  
$$
\mathcal{F}:  S \rightarrow 2^S
$$ 
using $F$, construct a retract
$$
R:  \mathbb{B}^2  \rightarrow S.
$$
By definition, a retract has the property that $R(v)=v$ for all $v \in S \subset \mathbb{B}^2$. Note that  
$$
\mathcal{F}(v)=F(R(v))
$$
so that the fixed points of $\mathcal{F}$ lie in $S$. 

\section{Conclusions}
We develop a game-theoretic formulation for compiling adiabatic quantum programs and point out the overlap with the minor graph embedding problem, which requires the embedded graph to satisfy the connectivity of a problem graph $\Gamma_P$ under the action of edge contraction. Our modification of the network creation game emphasizes that each player must now connect only to a subset of other players (nodes).
 
The problem of finding minor graph embeddings is discussed extensively in \cite{Choi1, Choi2}. One of the current leading methods for finding minor graph embeddings is the algorithm by Cai {\it et al.} \cite{Cai} which is based on an adaptation of Dijkstra's algorithm to generate a sought-after embedded graph. This so-called CMR algorithm starts by assigning two random vertices in $\Gamma_P$ to random positions in $\Gamma_H$ and then finding the shortest path between them. This process repeats, and with each iteration, adds a new vertex from $\Gamma_P$ into $\Gamma_H$ and forms paths that satisfy the connectivity. Embedded vertices may also grow into paths, as needed, to satisfy the connectivity requirements. The algorithm is probabilistic as it depends on the starting vertices and the shortest paths created between pairs. Consequently, for a give $\Gamma_H$ that is nearly the size of the input graph $\Gamma_P$, there is a significant probability that an embedding will not be found for arbitrary connectivity of $\Gamma_P$.
 
Our description of the compilation game is a first step toward presenting the CMR algorithm as a specific instance of the compilation game being played. By contrast, other current methods for constructing embeddings use knowledge of the connectivity of $\Gamma_H$ to synthesize $\Gamma_P$, though even these can certainly be related to the compilation game introduced here. 

A future direction of this work would be to consider if there exist alternative strategies in the compilation game that could increase the likelihood of finding the Nash equilibrium, by using for example the knowledge of the hardware connectivity.

\section{Acknowledgments}
This manuscript has been authored by UT-Battelle, LLC under Contract No. DE-AC05-00OR22725 with the U.S. Department of Energy. The United States Government retains and the publisher, by accepting the article for publication, acknowledges that the United States Government retains a non-exclusive, paid-up, irrevocable, world-wide license to publish or reproduce the published form of this manuscript, or allow others to do so, for United States Government purposes. The Department of Energy will provide public access to these results of federally sponsored research in accordance with the DOE Public Access Plan (http://energy.gov/downloads/doe-public-access-plan).

\end{document}